\documentclass[graybox]{svmult}

\usepackage{type1cm}        
\usepackage{makeidx}         
\usepackage{graphicx}        
\usepackage{multicol}        
\usepackage[bottom]{footmisc}

\usepackage{newtxtext}       %

\usepackage[hyperfootnotes=false]{hyperref} 

\usepackage{cite}

\DeclareSymbolFont{AMSa}{U}{msa}{m}{n}
\DeclareMathSymbol{\lesssim}      {\mathrel}{AMSa}{"2E}

\makeindex             

        \usepackage[normalem]{ulem}       
      \usepackage{xcolor}

\usepackage{graphicx}
\usepackage[figuresleft]{rotating}

\usepackage[greek,british]{babel}

\usepackage{graphicx}
\usepackage{dcolumn}
\usepackage{bm}

\usepackage{latexsym,array,theorem,mathrsfs,subfigure,bm,float}
\usepackage{stackengine}
\usepackage{amsmath}
\usepackage{amssymb}
\usepackage{xcolor}
\usepackage{mathtools}
\usepackage{tensor}
\usepackage[T1]{fontenc}
\usepackage{graphicx}
\usepackage[utf8]{inputenc}

\newcommand{\be}{\begin{eqnarray}}
\newcommand{\ee}{\end{eqnarray}}
\newcommand{\bea}{\begin{eqnarray}}
\newcommand{\eea}{\end{eqnarray}}

\newcommand{\A}{{\rm a}}

\newcommand{\nn}{\nonumber}

\newcommand{\To}{\Rightarrow}

\newcommand{\mmu}{\mu}

\newcommand{\kkappa}{\alpha}

\DeclareMathAlphabet\mathbfcal{OMS}{cmsy}{b}{n}

\begin{document}

\title*{Chaos in Horndeski cosmologies}

\author{Mikhail~S.~Volkov\orcidID{0000-0002-0843-4917}}

\institute{Institut Denis Poisson, UMR - CNRS 7013, Universit\'{e} de Tours, Parc de Grandmont, \\
37200 Tours, France;~~
\email{mvolkov@univ-tours.fr}\\
\\
Contribution to the book ``Open Issues in Gravitation and Cosmology - Original Contributions, Essays and Recollections in Honour of Alexei Starobinsky'', to be published by Springer, edited by Andrei Barvinsky and Alexander Kamenshchik.
}

\maketitle

 \abstract{
 We analyze how a scalar field can affect the chaotic behaviour of homogeneous and isotropic Bianchi IX cosmologies. 
 It is known that a massless, minimally coupled scalar field removes the chaos. However, in more general Horndeski 
 theories, the situation is more complex. We find that in shift-symmetric $K$-essence theories, chaos persists 
 if the scalar field contribution to the initial-value constraint is {\it subleading} compared to that of the anisotropies. 
 In this case, solutions oscillate as they approach the singularity, just as in the vacuum case, and a similar 
 behaviour is found when a non-minimal coupling is included.
If the scalar field contribution is not subleading, then chaos is removed and the singularity is approached smoothly. 
\indent
An unusual and entirely new result appears when changing the sign in front of the scalar kinetic term, yielding 
the theory of a phantom scalar. If the scalar field is subleading, then solutions remain chaotic and oscillate 
when approaching  the singularity, as before. However, if the scalar field is not subleading, solutions are also chaotic,
 but the spacetime singularity disappears, and the universe behaves as an apparently infinite sequence of anisotropic 
 bounces. The spatial volume then oscillates within finite bounds, never reaching zero, while the amplitudes and 
 positions of these oscillations appear completely random. To the best of our knowledge, this type of chaos 
 has never been described.
}

\section{Introduction}

I had the remarkable opportunity to work with Alexei Starobinsky during the last few years of his life, 
when we were members of a joint research project devoted to cosmologies in Horndeski theory. 
Together with other colleagues, we published three papers, starting from the homogeneous 
and isotropic case \cite{Starobinsky:2016kua} and then moving on to anisotropic models 
\cite{Starobinsky:2019xdp,Galeev:2021xit}. We obtained an interesting result: if the scalar 
field is non-minimally coupled, then in the Bianchi I case the anisotropies are suppressed 
(screened) and approach zero at the singularity instead of being amplified \cite{Starobinsky:2019xdp}.

However, including spatial curvature removes the screening \cite{Starobinsky:2019xdp}. 
The anisotropies then grow toward the singularity, and the Bianchi IX cosmology with 
a Horndeski scalar exhibits an oscillatory behaviour similar to the ``cosmological billiard'' of 
Belinskii-Khalatnikov-Lifshitz (BKL) \cite{Belinskii:1972} or the ``mixmaster universe'' of Misner 
\cite{Misner:1969hg} (see \cite{Goldstein:2025qxq} for a recent discussion). 
The billiard was originally discovered in the vacuum theory and is known to be chaotic, 
but similar behaviour persists when matter is added, 
provided that its energy density does not grow too rapidly near the singularity.

At the same time, it is known that adding a massless, minimally coupled scalar field suppresses the 
oscillations: the singularity is then approached smoothly, and the chaos disappears \cite{Belinskii:1973}. 
However, chaos persists for solutions with a non-minimally coupled scalar studied  in 
\cite{Starobinsky:2019xdp}. Since in both cases the scalar is described by Horndeski models, one may 
naturally ask when  the Horndeski theory admits chaotic solutions and when it does not.

I had intended to propose this as the next research project to Alexei Starobinsky, 
but we never had the opportunity to discuss it together. The project therefore remained only an idea, 
and it is only now that I am able to report some progress in this direction.

To clarify the issue, we shall consider below anisotropic cosmologies with a scalar field described 
by subclasses of Horndeski theory. In most cases, we examine the simplest shift-symmetric 
$K$-essence model. A natural conjecture is that chaos should persist if the scalar field contribution 
to the initial-value constraint is subleading compared to the anisotropy contribution, 
which scales as $\propto 1/\A^6$. For a minimally coupled massless scalar, its contribution 
also scales as $\propto 1/\A^6$, and therefore it is not subleading.

To test this conjecture, we generate numerical solutions and check whether they oscillate as they 
approach the singularity. Of course, this does not constitute a 
rigorous proof that the solutions are truly chaotic or oscillate infinitely many times, 
since numerical methods can only produce a finite number of 
oscillations before numerical errors grow rapidly. However, in some cases the solution profiles closely 
resemble those of the vacuum case, showing intervals during which one of the three scale factors is 
monotonic while the other two oscillate. In other cases, the oscillations cease quickly and all three scale 
factors become monotonic.

We therefore assume that a solution is chaotic if the numerical evolution displays oscillations that appear 
not to terminate. This is, of course, a very simple and naive definition of chaos, 
and we do not attempt to 
address numerous  subtle aspects of chaotic dynamics (some of which are discussed in 
\cite{Kamenshchik:1998ix}). Nevertheless, we verify that even an arbitrarily small change in the initial 
conditions completely reshuffles the positions and amplitudes of the oscillation peaks. This sensitivity 
implies positive Lyapunov exponents, confirming chaotic behaviour.

We find that the above conjecture holds for standard $K$-essence with a positive kinetic energy 
of the scalar field. The scalar energy scales as $1/\A^\gamma$, where $\gamma$ is a parameter 
of the theory. For $\gamma < 6$, the scalar contribution is subleading compared to the anisotropy effect, 
and the solutions oscillate as they approach the singularity. For $\gamma \geq 6$, the scalar effect 
is not subleading, the oscillations cease, 
and the singularity is approached smoothly.

However, a completely new and unexpected feature arises when changing the sign of the scalar kinetic 
term. This yields the theory of a phantom scalar field whose energy scales as $-1/\A^\gamma$. 
For $\gamma < 6$, the solutions oscillate as they approach the singularity and display chaos of the usual 
BKL type. Surprisingly, for $\gamma \geq 6$, they also oscillate and appear chaotic, but the spacetime 
singularity disappears.

The solution then becomes globally regular and exhibits a sequence of anisotropic bounces with 
chaotically distributed minimal and maximal sizes. Within each bounce, the three anisotropy amplitudes 
undergo a finite number of oscillations, but the number of bounces is seemingly infinite, and the universe 
extends indefinitely into both the past and future --  it is eternal (see \cite{Easson:2024fzn} 
for other examples of eternal cosmologies). 
Therefore, our conjecture fails in the case of a phantom scalar, but a new form of chaotic behaviour emerges that,
 to the best of our knowledge, has never been described.

Finally, we analyze the theory with a non-minimal coupling considered in \cite{Starobinsky:2019xdp}. 
The solutions approach the singularity and oscillate as in the BKL case, and we verify that the scalar 
contribution is subleading, which once again supports the conjecture.

In our analysis, we combine analytical and numerical methods. Of course, one cannot describe 
Bianchi IX solutions fully analytically, but they can be piecewise approximated by 
the ``Kasner epochs'' described by 
Bianchi I solutions.
In the vacuum case, the Bianchi I solution (Kasner metric) is characterized by three exponents 
$p_1, p_2, p_3$, one of which is always negative, so that one of the three scale factors
$\A_k \propto t^{p_k}$ grows toward the singularity. This feature also appears in vacuum Bianchi IX 
solutions: the scale factors oscillate, and there is always one that grows toward the singularity.

When a massless scalar is included, one of the three exponents $p_k$ in the Bianchi~I case can be 
negative, but it also becomes possible for all three to be positive. It is this latter property that halts the 
oscillations in the Bianchi IX case. The values of $p_k$ change after each oscillation cycle, and the 
oscillations continue as long as one of the three $p_k$ is negative. However, sooner or later the 
system inevitably reaches a state in which all three $p_k$ are positive, and then all three scale factors 
become monotonic, causing the oscillations to stop \cite{Belinskii:1973}. In what follows, we review this 
property.

We then consider the Bianchi I solutions with a phantom scalar field and find that among the three 
exponents $p_k$, {\it either one or two} can be negative, and all three can never be positive. This implies 
that the Bianchi IX solutions cannot become smooth, since all three scale factors cannot be monotonic, 
and either one or two of them grow toward the past. These properties are confirmed by the numerical 
solutions.

\section{The theory}

We consider the  theory 
\be                             \label{0}
S=\frac12\int\left(\mmu\, R
+\epsilon \,X^n
-\kkappa\, G_{\mu\nu}\,\partial^\mu\phi\,\partial^\nu\phi
\right)\sqrt{-g}\, d^4x
\equiv\frac12 \int L\,d^4x\,.
\ee
Here  $R$ and $G_{\mu\nu}$ are the Ricci scalar and Einstein tensor, and 
$
X=-\partial^\mu \phi\, \partial_\mu\phi, 
$
while $\mu$ is related to the Planck mass. 
The scalar field dynamics is determined by the $K$-essence term $\epsilon X^n$ and by the term 
proportional to $\kkappa$, which provides the non-minimal coupling. 
We refer to $n$ as the $K$-essence index.

The parameter $\epsilon$ takes the values $\epsilon = +1$, 
corresponding to an ordinary scalar field, or $\epsilon = -1$, 
corresponding to a phantom scalar. For $n = 2$, 
the theory \eqref{0} reduces to the one studied in \cite{Starobinsky:2019xdp} 
(up to a cosmological term, which we do not include).

We shall assume the scalar field to depend only on time, $\phi=\phi(t)$, 
and consider homogeneous and anisotropic metrics of the form 
\be                        \label{maa}
ds^2=-N^2(t)\,dt^2
+\left.\left.\frac14\,\right[\A_1^2(t)\,\omega^1\otimes \omega^1
+\A_2^2(t)\,\omega^2\otimes \omega^2
+\A_3^2(t)\,\omega^3\otimes \omega^3\right].
\ee
Here $\omega^a = \omega^a_{\,k}\, dx^k$ are spatial one-forms. 
In the simplest Bianchi I case, one has $\omega^a = dx^a$, 
while in the Bianchi IX case $\omega^a$ are invariant forms on $S^3$ satisfying
$
d\omega^a+\epsilon_{abc}\,\omega^b\wedge \omega^c=0$.
The factor $1/4$ in \eqref{maa} is introduced to match the convention used in \cite{Starobinsky:2019xdp}.

Injecting this into \eqref{0} yields 
\be         \label{LL}
L=\mu N\left(-T+V \right) -
\alpha\,\frac{\dot{\phi}^2}{2N} \left(T+V \right)
+
\epsilon \left(\frac{\dot{\phi}}{N}\right)^n
N\, \A_1 \A_2 \A_3 \,,
\ee
where 
\be
T=\frac{1}{4N^2}\left(\A_1\,\dot{\A}_2\,\dot{\A}_3+\A_2\,\dot{\A}_1\,\dot{\A}_3+\A_3\,\dot{\A}_1\,\dot{\A}_2\right)\equiv \frac{3}{4}\,\A_1\A_2\A_3\times y\,,
\ee
and in the Bianchi IX case one has 
\be
V&=&-\frac{(\A_1+\A_2+\A_3)(\A_1+\A_2-\A_3)(\A_1-\A_2+\A_3)(\A_1-\A_2-\A_3)}{4\A_1\A_2\A_3} \nn \\
&&\equiv \frac34\, (\A_2\A_2\A_3)^{1/3}\times {\cal K},~~~~~ 
\ee
whereas  in the Bianchi I case $V=0$. 
Setting 
\be
\A_1=\A\, e^{\beta_{+}+\sqrt{3}\beta_{-}},~~
\A_2=\A\, e^{\beta_{+}-\sqrt{3}\beta_{-}},~~
\A_3=\A\, e^{-2\beta_{+}}~~~~\Rightarrow~~~\A=\A_1 \A_2 \A_3\,,
\ee
the Lagrangian in \eqref{LL} assumes the form 
\be
8L=6\mmu\A^3 N\left(\frac{{\cal K}}{\A^2}-y  \right)
-\frac{3\kkappa\A^3}{N}\,\dot{\phi}^2\left(y+\frac{{\cal K}}{\A^2} \right)
+
\epsilon \left(\frac{\dot{\phi}}{N}\right)^n
N\A^3\,,
\ee
with 
\be                \label{K}
y&=&\left.\left.\frac{1}{N^2}\right(\frac{\dot{\A}^2}{\A^2}-\dot{\beta}_{+}^2-\dot{\beta}_{-}^2\right),\nn \\
{\cal K}&=&-\frac13\,e^{-8\beta_{+}}
\left(4e^{6\beta_{+}}\cosh^2(\sqrt{3}\beta_{-})-1\right)
\left(4e^{6\beta_{+}}\sinh^2(\sqrt{3}\beta_{-})-1\right).
\ee
Varying the Lagrangian then  gives 
equations 
\bea             \label{K1}
3\mmu\left(y+\frac{\cal K}{\A^2}\right)
+\frac{3}{2}\,\kkappa\,\psi^2\left(3y+\frac{\cal K}{\A^2}\right)- 
\epsilon\,\frac{n-1}{2}\,\psi^{~}
\equiv {\cal C}=0,           \\                 
\left.\left.\frac{1}{\A^2N}\right(\frac{\sigma_{+}\, \A\dot{\A}}{N}\right)^{\mbox{.}}
=\frac{\sigma_{+}}{2N^2}\left(\frac{\dot{a}^2}{a^2}
-3\dot{\beta}^2_{+}
-3\dot{\beta}^2_{-}\right)
-\frac{\sigma_{-}\,{\cal K}}{2\A^2}
-\frac{\epsilon}{2}\,\psi^n
,           \label{K2}    \\
\frac{1}{N}\left(\frac{\sigma_{+}\,\A^3}{N} \dot{\beta}_\pm \right)^{\mbox{.}}=
\frac{\sigma_{-}\,\A}{2}\,\frac{\partial{\cal K}}{\partial \beta_\pm},        \label{K3}   \\
\A^3\left(
-\frac{6\kkappa}{n\epsilon }\,\left(y+\frac{\cal K}{\A^2}\right)+
\psi^{n-2}\right)\psi=\sqrt{6\mu}\,C,    \label{K4} 
\eea
with 
\be
\psi=\frac{\dot{\phi}}{N},~~~~\sigma_\pm =2\mu\pm \alpha\psi^2. 
\ee
Eq.\eqref{K1} is the constraint ${\cal C}=0$, which restricts  the initial values.
This is a first integral whose 
derivative vanishes due to the remaining Eqs.\eqref{K2}--\eqref{K4}, $\dot{\cal C}=0$, so it is 
sufficient to impose this constraint only at the initial moment of time. 
Eq.\eqref{K4} is also a first integral, arising from the 
invariance of the theory under shifts 
\be
\phi\to\phi+const,
\ee
which implies conservation of the scalar charge --
the integration constant $C$ in \eqref{K4}. 

 The function $N$ is a gauge parameter that enters the equations only through the 
combination
\be
\frac{1}{N}\,\frac{d}{dt}\equiv \frac{d}{d x},
\ee
so different choices of $N$ correspond to different time coordinates.
In our numerical analysis,  
we use  two gauges, $N=1$ and $N=\A^3$, and it turns out that 
functions such as  $\beta_\pm(a)$ or $\psi(a)$ do not depend of the gauge choice. 
In some cases, the functions are shown against the physical time, defined in a 
gauge-invariant way as
\be
t=\int_0^x\frac{d x}{N}\,.
\ee

To solve the equations numerically, we consider the second-order equations \eqref{K2},\eqref{K3}, 
together with  the derivative of \eqref{K4}. Resolving  with respect 
to the highest derivatives yields the system 
\be                     \label{eqs}
\ddot{\A}&=&A(N,a,\dot{a},\beta_\pm,\dot{\beta}_\pm,\psi),~~~~\nn \\
\ddot{\beta}_\pm&=&B_\pm(N,a,\dot{a},\beta_\pm,\dot{\beta}_\pm,\psi),~~~~\nn \\
\dot{\psi}&=&F(N,a,\dot{a},\beta_\pm,\dot{\beta}_\pm,\psi).
\ee
Setting, for example, $N=1$, and choosing the initial values at $t=0$, 
\be
\A(0),~~\dot{\A}(0),~~\beta_\pm(0),~~\dot{\beta}_\pm(0),~~\psi(0),
\ee
which mush satisfy the constraint \eqref{K1}, the equations \eqref{eqs} can be integrated 
numerically toward  either negative of positive values of $t$. During the integration, the 
constraint and the scalar charge should be monitored to ensure  that 
\be
{\cal C}=0,~~~~~~~C=const.
\ee
The integration continues as long as the value of ${\cal C}$ remains small, and stops when 
it starts to grow rapidly.  

Before integrating the equations numerically, some preliminary analysis is required. 
All formulas below correspond to the gauge choice  $N=1$ .

\section{ Bianchi I, vacuum solution}

It is consistent to set $\psi=C=0$ in the equations.  In the Bianchi I case, when 
${\cal K}=0$,  the equations yield 
\be                    \label{K0}
 \frac{\dot{\A}^2}{\A^2}=\dot{\beta}_{+}^2+\dot{\beta}_{-}^2,~~~~~
 \dot{\beta}_{\pm}=\frac{{\cal B}_{\pm}}{ \A^3 },
 \ee
 with ${\cal B}_\pm$ being integration constants. Denoting 
 \be
 {\cal B}=\sqrt{{\cal B}_{+}^2+{\cal B}_{-}^2}
 \ee
 and integrating, one obtains 
 \be
 \A^3=3{\cal B}\, t,~~~~~
 \beta_\pm =\frac{{\cal B}_\pm }{3{\cal B}}\,\ln(t),
 \ee
 which corresponds to the Kasner metric, for which   $\A_k\propto t^{p_k}$ with 
\be
p_1=\frac13+\frac{{\cal B}_{+}+\sqrt{3}\,{\cal B}_{-}}{3{\cal B}},~~~
p_2=\frac13+\frac{{\cal B}_{+}-\sqrt{3}\,{\cal B}_{-}}{3{\cal B}},~~~
p_3=\frac13-\frac{2{\cal B}_{+}}{3{\cal B}}.
\ee 
One has 
\be             \label{pp}
p_1+p_2+p_3=p_1^2+p_2^2+p_3^2=1. 
\ee
These relations imply that one of the three coefficients $p_{k}$ is negative while the other two 
are positive. Therefore, one of the three metric functions  $\A_k(t)$ grows
as $t\to 0$, while the other two decrease. 
All possible values of $p_k$ can be parameterized as follows \cite{Belinskii:1972}:
\be               \label{pu}
p_1(u)=\frac{-u}{1+u+u^2},~~~~
p_2(u)=\frac{1+u}{1+u+u^2},~~~~
p_3(u)=\frac{u(1+u)}{1+u+u^2},~~~~
\ee
where $0\leq u\leq 1$. 

\section{ Bianchi I with a massless scalar field}

Let us include the scalar field, assuming that it is minimally coupled (hence $\alpha=0$)
and that the $K$-essence index $n=2$. 
The equations reduce to 
\be                    \label{K00}
 \frac{\dot{\A}^2}{\A^2}=\dot{\beta}_{+}^2+\dot{\beta}_{-}^2+\frac{\epsilon}{6\mu}\,\psi^2,~~~~~
 \dot{\beta}_{\pm}=\frac{{\cal B}_{\pm}}{ \A^3 },~~~~~~~\psi=\frac{\sqrt{6\mu}\,C}{\A^3},
 \ee
 and therefore
 \be            \label{aa}
 \frac{\dot{\A}^2}{\A^2}-\frac{{\cal B}^2+\epsilon \,C^2}{\A^6}=0.
 \ee
 If we interpret this as a sum of kinetic and  potential terms, then it follows that the anisotropy 
 contribution to the potential, $-{\cal B}^2/\A^6$, provides an attraction toward singularity at $\A=0$. 
 The scalar contribution  $-\epsilon C^2/\A^6$ is also attractive if $\epsilon=1$, but 
 becomes repulsive when  $\epsilon=-1$. 
 
 Integrating, one obtains 
  \be                               \label{A3}
 \A^3=3\sqrt{{\cal B}^2+\epsilon\, C^2}\,t=\frac{\sqrt{6}\,C}{q}\, t,~~~
 \dot{\beta}_\pm =\frac{q}{\sqrt{6}\,C}\,\frac{{\cal B}_\pm}{t},~~~
 \psi=\frac{\sqrt{\mu}\,q}{t},~~~\A_k\propto t^{p_k},~~~~
 \ee
 where the parameter $q$ is related to the scalar charge $C$,
 \be              \label{q}
 q=\frac{\sqrt{2/3}\,C}{\sqrt{ {\cal B}^2+\epsilon\,C^2}}~~~~\Leftrightarrow~~~~
 C=\frac{q\, {\cal B} }{\sqrt{2/3-\epsilon\, q^2} }. 
 \ee
 One has 
 \be                   \label{pb}
 p_1&=&\frac13+\frac{q}{\sqrt{6}\,C}\,({\cal B}_{+}+\sqrt{3}{\cal B}_{-}),~~~
 p_2=\frac13+\frac{q}{\sqrt{6}\,C}\,({\cal B}_{+}-\sqrt{3}{\cal B}_{-}),~~~\nn \\
 p_3&=&\frac13-\frac{2q}{\sqrt{6}\,C}\,{\cal B}_{+}, ~~~~~~~~~
  \ee
 which implies 
  \be                  \label{pp1}
p_1+p_2+p_3=1,~~~~p_1^2+p_2^2+p_3^2=1-\epsilon\, q^2,
\ee
where the principal difference compared  with \eqref{pp} is the term $\epsilon\, q^2$ 
appearing on the right-hand side. 
In this case there is also a parametric representation similar to \eqref{pu} \cite{Belinskii:1973},
\be            \label{pu1}
p_1(u)&=&\frac{-u}{1+u+u^2},~~~~~~~ \nn \\
p_2(u)&=&\frac{1+u}{1+u+u^2}\,\left(u-\frac{u-1}{2}(1-\sqrt{1-B} ) \right), \nn \\
p_3(u)&=&\frac{1+u}{1+u+u^2}\,\left(1+\frac{u-1}{2}(1-\sqrt{1-B} ) \right), \nn \\
B&=&2\epsilon\,q^2\, \frac{(1+u+u^2)^2}{(u^2-1)^2}.
\ee
One has 
\be
p_1\left(\frac{1}{u}\right)=p_1(u),~~~~p_2\left(\frac{1}{u}\right)=p_3(u),
~~~~~~p_3\left(\frac{1}{u}\right)=p_2(u),
\ee
therefore one can assume that $u\in [-1,1]$. 
Eq.\eqref{pu1} reduces to \eqref{pu} if the scalar charge $C$ vanishes. 
Notice also that choosing a value of $u$ 
in \eqref{pu1} determines the anisotropy parameters ${\cal B}_\pm$ in \eqref{pb}
as follows:
\be
{\cal B}_{+}=\frac{C}{\sqrt{2}q}(p_1(u)-p_2(u)),~~~~
{\cal B}_{-}=\frac{C}{\sqrt{6}q}(1-3p_3(u)).
\ee
Suppose now that $C\neq 0$  and consider separately the two cases $\epsilon=+1$
and $\epsilon=-1$.

\subsection{Bianchi I with an ordinary scalar field, $\epsilon=1$}
\begin{figure}[h]
\hbox to \linewidth{ \hss

				\resizebox{6.3cm}{5cm}{\includegraphics{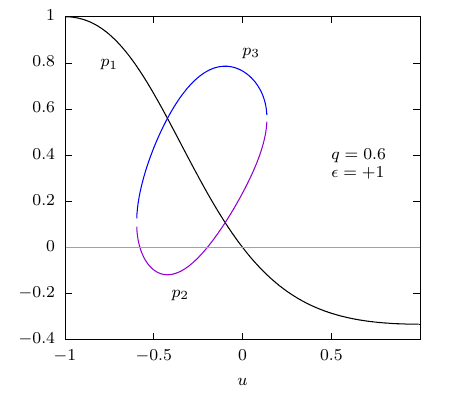}}
				\resizebox{6.3cm}{5cm}{\includegraphics{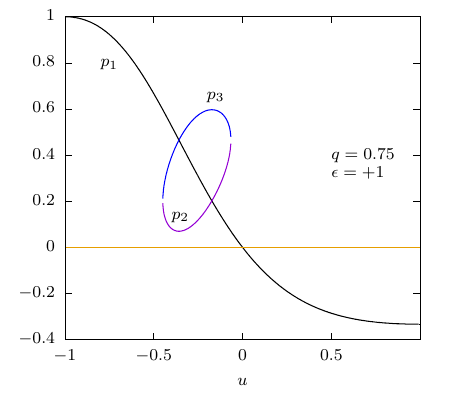}}	
\hspace{1mm}
\hss}
\caption{Left: Coefficients $p_k(u)$ for $q=0.6$, $\epsilon=+1$. All three $p_k(u)$'s
exist for $u\in (-0.596,0)$, and  they  are all 
positive for $u\in (-0.596,-0.578)$ and for $u\in (-0.198,0)$. Right: 
Coefficients for $q=0.75$; they exist for $u\in (-0.449,-0.065)$
and are all positive. 
In the limit $q\to \sqrt{2/3}\approx 0.816$, the oval formed by $p_2$ and $p_3$ 
shrinks to the point at $u=-2+\sqrt{3}\approx -0.267$, and then $p_1=p_2=p_3=1/3$. 
}
 \label{Fig2}
\end{figure}

In this case, 
as shown by \eqref{q}, 
when the scalar charge varies in the interval
$C\in [0,\infty)$, one has $q^2\in [0,2/3)$. 
The values of $p_k(u)$ for $q\neq 0$ are shown in Fig.\ref{Fig2}, 
and one can see that all three of them  exist only if the range of $u$ is 
restricted to the interior of the oval region delimited by $p_2(u)$ and $p_3(u)$,
hence $u\in [u_{\rm min}(q),u_{\rm max}(q)]$. 
This interval of allowed values  shrinks as $q$ grows, so that for $q=0$ 
one has $u\in[-1,1]$; if $q=1/2$ then $u\in[-0.67,0.25]$; and when 
$q\to \sqrt{2/3}$ 
the interval shrinks to a single point $u=-2+\sqrt{3}\approx -0.267$. 

When $q$ is small, then for most values of $u$ from the interval 
$[u_{\rm min}(q),u_{\rm max}(q)]$ one of the three $p_k(u)$ is negative, 
while the other  two are  positive. However, there are always values of $u$ 
for which all  three $p_k$'s are positive, as seen in Fig.\ref{Fig2} (left panel). 
For $q>1/\sqrt{2}$, all three $p_k$'s are always positive,
as shown in Fig.\ref{Fig2} (right panel). If all three $p_k$'s are positive,
then all three scale factors in the metric 
\be
\A_k\propto t^{p_k}
\ee
approach zero for $t\to 0$. As we shall see below, this eliminates the chaotic solutions 
in the Bianchi IX case. 
Notice also that the scale factor in \eqref{A3}, 
\be                      \label{A3aa}
\A^3=3 \sqrt{{\cal B}^2+C^2}\, ,
\ee
admits the isotropic limit where ${\cal B} \to 0$. 

\subsection{Bianchi I with a phantom scalar field, $\epsilon=-1$}
\begin{figure}[h]
\hbox to \linewidth{ \hss

				\resizebox{6.3cm}{5cm}{\includegraphics{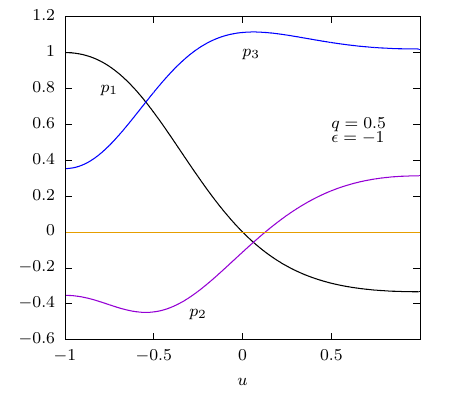}}
				\resizebox{6.3cm}{5cm}{\includegraphics{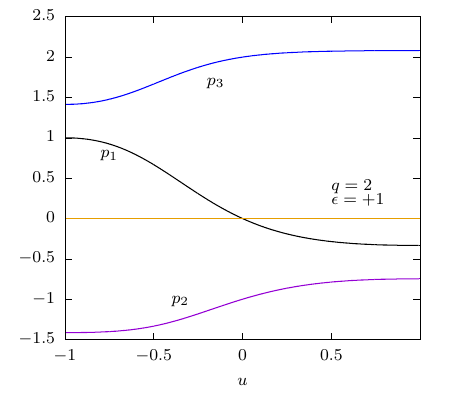}}	
\hspace{1mm}
\hss}
\caption{Coefficients $p_k(u)$ for $\epsilon=-1$ and for either $q=0.5$ (left) or
$q=2$ (right). They never all become positive, but two of them can be negative simultaneously.
}
 \label{Fig3}
\end{figure}

 The phantom scalar  provides a repulsion, and 
Eq.\eqref{aa} assumes the form 
 \be            \label{aaa}
 \frac{\dot{\A}^2}{\A^2}=\frac{{\cal B}^2- \,C^2}{\A^6},
 \ee
 whose solution is 
 \be                      \label{A3aB}
\A^3=3 \sqrt{{\cal B}^2-C^2}\, t\,.
\ee
The solution exists only if anisotropies do not vanish and 
the scalar charge 
does not exceed 
the maximal value, $|C|<{\cal B}$. When $|C|$ ranges 
within the interval $[0,{\cal B})$, one has $q^2\in [0,\infty)$. Values of $p_k(u)$ for different $q$ 
are shown in Fig.\ref{Fig3}, where we see that the interval of allowed $u$ 
is always $[-1,1]$. In addition, we see that {\it either one or two}
of the three $p_k(u)$'s 
are negative, and all three can never be positive. This has important consequences 
in the Bianchi IX case. 

Summarizing, the solution is similar to the Kasner metric, with 
\be
\A_k\propto t^{p_k},
\ee
where either one or two of the three $\A_k$'s grow toward  the past.

\section{ Bianchi IX, vacuum solutions}

Let us examine how the Bianchi I  solutions are modified by the presence of the spatial curvature. 
If ${\cal K}\neq 0$ but $\psi=0$, then the constraint equation \eqref{K1}  becomes
\be                   \label{cons}
 \frac{\dot{\A}^2}{\A^2}+\frac{\cal K}{\A^2}=\dot{\beta}_{+}^2+\dot{\beta}_{-}^2\,.
 \ee
 The function ${\cal K}(\beta_{+},\beta_{-})$ defined in \eqref{K} takes values in the range
 \be
 {\cal K}(\beta_{+},\beta_{-})\in(-\infty,1]~~~~\text{with}~~~1={\cal K}(0,0). 
 \ee
 We assume that at the initial moment $t=0$ 
 one has 
 \be              \label{ini1}
 \A(0)=1,~~~\beta_\pm(0)=0,~~~~
 \dot{\beta}_{+}(0)={\cal B}\,\sin\gamma,~~~~\dot{\beta}_{-}(0)={\cal B}\cos\gamma,
 \ee
 where ${\cal B}$ and $\gamma$ are constants. 
 The constraint \eqref{cons} is satisfied if
 \be            \label{ini2}
 \dot{\A}(0)=0,~~~~~{\cal B}=1.
 \ee
  By choosing $\gamma$, we use \eqref{ini1} and \eqref{ini2} as initial 
  conditions for second-order equations \eqref{K2} and \eqref{K3}. Their solution 
  determines $\A(t)$ and $\beta_\pm(t)$, and   the three amplitudes 
  $\A_k(t)$. 
  
  Starting at $t=0$, we integrate into the $t<0$ region. The 
  system approaches a singularity such that the three amplitudes $\A_k$ 
  oscillate, but  one of them always grows  toward the past, 
   while the product $V=\A_1 \A_2 \A_3$ decreases monotonically and approaches zero.
  The same patterns is observed when we 
  integrate toward positive values of $t$, even if we choose in 
  \eqref{ini2} ${\cal B}>1$ and $\dot{\A}(0)\neq 0$. Therefore, the spacetime contains
  singularities both in the past and in the future. 
  
\begin{figure}[h]
\hbox to \linewidth{ \hss

				\resizebox{6.3cm}{5cm}{\includegraphics{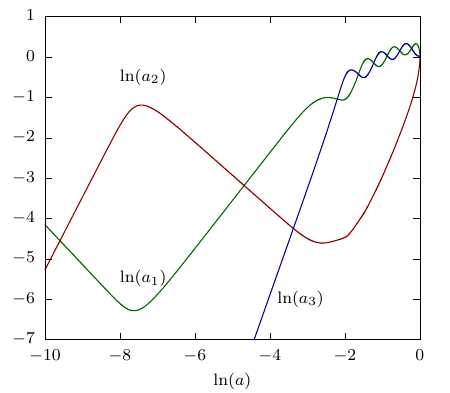}}	
				\resizebox{6.3cm}{5cm}{\includegraphics{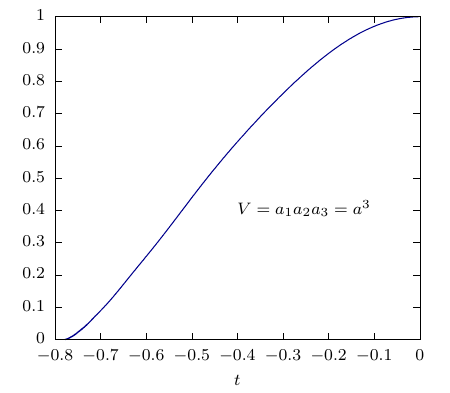}}

\hspace{1mm}
\hss}
\caption{The vacuum Bianchi IX solution for the initial data \eqref{ini1},\eqref{ini2} 
with $\gamma=3$. The three amplitudes $\A_k$ oscillate, exhibiting the characteristic ``billiard'' behaviour. 
Their product $V=\A_1\A_2\A_3=\A^3$ is monotonic and approaches zero at the singularity
}
 \label{Fig0}
\end{figure}

 We choose  the value $\gamma=3$ to illustrate the typical solution in Fig.\ref{Fig0}.
 The figure shows the characteristic ``billiard'' behaviour: 
  the singularity is approached through a sequence of oscillation cycles (Kasner epochs). 
  During each cycle, 
one has $\partial{\cal K}/\partial \beta_\pm\approx 0$, and 
the geometry is approximately described by the Kasner metric, with $\A_k\propto  t^p_k$
and $p_k=p_k(u)$ for some $u$. The term 
$\partial{\cal K}/\partial \beta_\pm$ becomes 
significant  only at the end of each cycle, where  $u$ changes, 
initiating the next oscillation cycle
with new values of $u$ and $p_k(u)$.

As seen in Fig.\ref{Fig0} (right panel), the scale factor $\A$ is a monotonic function of $t$,
so $\ln(\A)$ can be used as a convenient time coordinate (left panel).  
Within a given interval of $\ln(\A)$, one of the three amplitudes $\A_k$ has an exponent $p_k$ that remains 
positive, making this amplitude monotonic. For example, $\A_2$ is monotonic for $\ln(\A)\in (-2,0)$ 
in Fig.\ref{Fig0}.
The other two amplitudes exhibit several oscillations over this range, forming a ``braid'' pattern. 
After each oscillation, their exponents interchange signs, as illustrated by 
$\A_1$ and $\A_3$ for $\ln(\A)\in (-2,0)$. Outside this interval, one of the previously oscillating amplitudes 
becomes monotonic, e.g., $\A_3(t)$ for $\ln(\A)<-2$ in Fig.\ref{Fig0}, while the remaining two begin to oscillate.

The solutions are chaotic, and the changes of $u$ after each oscillation cycle follow a statistical pattern \cite{Belinskii:1972}.
Although the approach to the singularity involves infinitely many oscillations,
only a finite number of them can be captured in numerical simulations.

A similar oscillatory regime persists also  when matter is included,
provided its energy density grows more slowly than $1/\A^6$ as $\A\to 0$,
so that its contribution to the constraint \eqref{K0} remains subdominant 
compared to the anisotropies.
This applies, for instance, to dust or a perfect fluid.
However, the presence of a massless scalar field destroys the chaotic behaviour \cite{Belinskii:1973}.

\section{ Bianchi IX with a massless scalar field}

Including the massless scalar field, 
the initial value constraint \eqref{K1} becomes 
\be                    \label{K00o}
 \frac{\dot{\A}^2}{\A^2}+\frac{\cal K}{\A^2}=\dot{\beta}_{+}^2+\dot{\beta}_{-}^2+\frac{\epsilon}{6\mu}\,\psi^2. 
 \ee
 This relation fixes the initial conditions, which we choose in the form analogous to \eqref{ini1},\eqref{ini2}:
  \be              \label{ini3}
 \A(0)&=&1,~~~\beta_\pm(0)=0,~~~~
 \dot{\beta}_{+}(0)={\cal B}\,\sin\gamma,~~~~\dot{\beta}_{-}(0)={\cal B}\cos\gamma, \nn \\
 \dot{\A}(0)&=&0,~~~~~~~\psi=\sqrt{6\mu} C,~~~~~~{\cal B}=\sqrt{1-\epsilon\, C^2}, 
 \ee
 where $C$ is the scalar charge. 
 It turns out that solutions behave very differently depending on whether $\epsilon=+1$ or $\epsilon=-1$.

 \subsection{Bianchi IX with  an ordinary scalar, $\epsilon=1$}

 \begin{figure}[h]
\hbox to \linewidth{ \hss

	
				\resizebox{6.3cm}{5cm}{\includegraphics{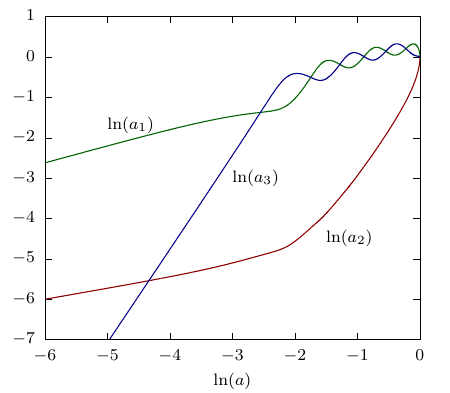}}
				\resizebox{6.3cm}{5cm}{\includegraphics{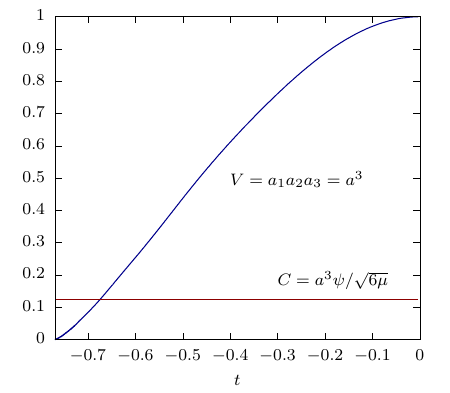}}	

\hspace{1mm}
\hss}
\caption{The Bianchi IX solution with a massless scalar ($\epsilon=1$), 
for the initial data \eqref{ini3} with $\gamma=3$ and $C=0.123$.
After a few oscillations, all three functions $\A_k$ become monotonic.
The volume $V=\A^3$ decreases to zero, while the scalar charge $C$ defined by \eqref{K4}
remains constant, implying $\psi\propto C/\A^3$.
}
 \label{Fig4a}
\end{figure}

In this case, the scalar charge satisfies $|C|\in[0,1]$. 
If  $|C|=1$, then  ${\cal B}=0$, 
and both the anisotropies and their first derivatives vanish at the initial time,
so they remain zero for all times.
The constraint equation 
\eqref{K00o} then reduces to 
\be                    \label{K001}
 \frac{\dot{\A}^2}{\A^2}+\frac{1}{\A^2}=\frac{1}{\A^6}, 
 \ee
 whose solution describes an isotropic universe that begins with unit size and contracts to zero volume.
 If instead one chooses $|C|<1$ in \eqref{ini3}, the solutions become anisotropic generalizations of this case:
 the universe undergoes oscillations while shrinking, but only a finite number of oscillation cycles occur.

 The typical solution is shown in Fig.\ref{Fig4a}, corresponding to $\gamma=3$ and $C=0.123$.
Near the initial time, for $\ln(\A)\in (-2,0)$, the solution resembles the vacuum case in Fig.\ref{Fig0} 
and exhibits oscillations.
These oscillations form Kasner-type cycles, during which one of the three exponents $p_k(u)$ is 
negative while the other two are positive.
After each cycle, the parameter $u$ changes, yielding new values $p_k(u)$ as defined in \eqref{pu1}.
If one of the exponents remains negative, the oscillations continue.

However, as shown in \cite{Belinskii:1973}, sooner  or later the system reaches a value of $u$ 
for which all $p_k(u)$ are positive,
so that all three metric functions $\A_k$ become monotonic.
At this point, the oscillations cease, and the singularity is approached smoothly.
Indeed, Fig.\ref{Fig4a} shows that all three $\A_k$ become monotonic for $\ln(\A)<-2$, and none of them grows
toward the pas. 

In summary, the inclusion of a minimally coupled scalar field permits only a 
finite number of oscillations near the singularity, thereby eliminating chaotic behaviour \cite{Belinskii:1973}. 
By contrast, the dynamics are entirely different in the presence of a minimally coupled phantom scalar field.

 \subsection{Bianchi IX with a phantom scalar, $\epsilon=-1$}
 
Recall that a phantom scalar field provides a repulsion from the singularity.
No isotropic limit is possible in this case, since achieving it would require satisfying an equation with a 
sign change on the right-hand side in \eqref{K001}, which is not possible.
 
\begin{figure}[h]
\hbox to \linewidth{ \hss

	
				\resizebox{6.3cm}{5cm}{\includegraphics{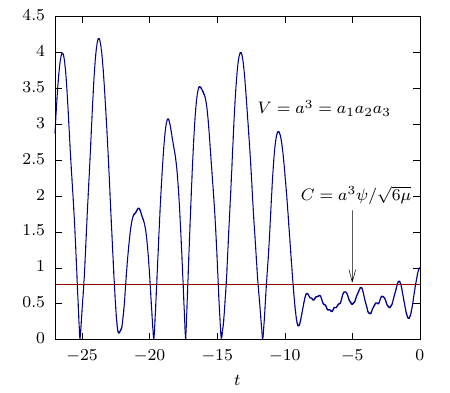}}
				\resizebox{6.3cm}{5cm}{\includegraphics{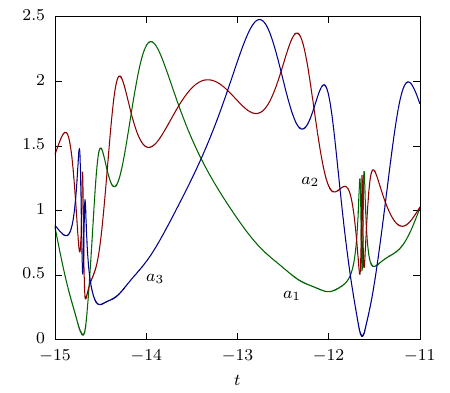}}	

\hspace{1mm}
\hss}
\caption{Bianchi~IX solution with a phantom scalar field ($\epsilon=-1$)  for the initial values
\eqref{ini3} with $\gamma=-2$ and $C=0.774$. Left: the spatial volume  $V(t)=\A_1\A_2\A_3$ 
and scalar charge $C=\A^3\psi/\sqrt{6\mu}$. Right: the scalar factors $\A_1(t)$, $\A_2(t)$, $\A_3(t)$ 
in the interval $t\in [-15,-11]$. 
}
 \label{Fig4b}
\end{figure}

 For the anisotropic solutions, the scalar charge 
 can assume any value, $|C|\in [0,\infty)$, while ${\cal B}=\sqrt{1+C^2}$. 
 Starting from the initial values \eqref{ini3} with $\epsilon=-1$, $C=0.774$,  and $\gamma=-2$, 
 and integrating toward negative $t$, one obtains the behaviour shown in Fig.\ref{Fig4b}.
 
The spatial  volume $V=\A_1 \A_2 \A_3\equiv \A^3$ oscillates 
chaotically, sometimes approaching zero but never vanishing, and repeatedly bouncing back.

The function $V(t)$ in the left panel in Fig.\ref{Fig4b} exhibits multiple minima, two of which occur  in 
the interval $t\in [-15,-11]$. The right panel of Fig.\ref{Fig4b} shows a zoom of this interval, 
revealing  that 
all three amplitudes $\A_k(t)$ oscillate, with particularly dense oscillations 
around $t=-14.7$ and around $t=-11.6$. 

\begin{figure}
\hbox to \linewidth{ \hss

	
				\resizebox{6.3cm}{5cm}{\includegraphics{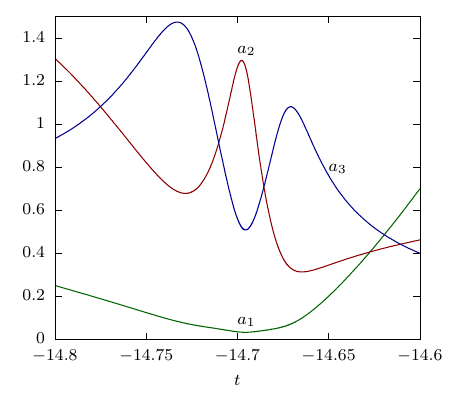}}
				\resizebox{6.3cm}{5cm}{\includegraphics{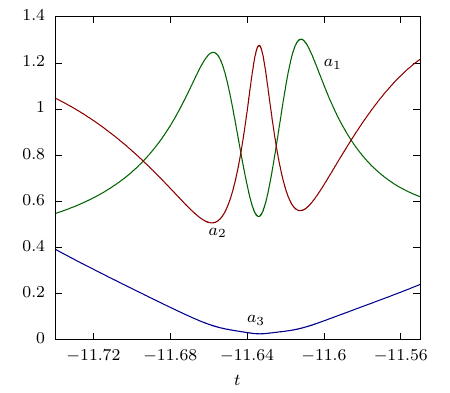}}	

\hspace{1mm}
\hss}
\caption{Same as the right panel of Fig.\ref{Fig4b}, zoomed around the two regions 
where the oscillations become particularly dense.
}
 \label{Fig4c}
\end{figure}

Zooming further into these regions of dense oscillations, the left panel of
Fig.\ref{Fig4c} shows that $\A_1(t)$ passes through a non-zero minimum at 
$t\approx -11.69$, while $\A_2(t)$ and $\A_3(t)$ oscillate several times around each   other. 
Similarly, the  right panel of Fig.\ref{Fig4c} shows that $\A_3(t)$ 
reaches a non-zero minimum at $t\approx -11.63$, with $\A_1(t)$ and $\A_2(t)$ 
oscillating around one another.

In the left panel in Fig.\ref{Fig4c}, to the right of the minimum of $\A_1$,
one of the three amplitudes grows toward the past while the  two others decrease,
so that among the three Kasner exponents $p_k$, one is negative and two are positive. 
Conversely, to the left of the minimum of $\A_1$,  
{\it two} of the three amplitudes grow toward the past and the third decreases, so that 
among the three Kasner exponents $p_k$, {\it two} are negative and one is positive.
This behaviour is consistent with the previous analysis of the allowed values of $p_k$  in the Bianchi~I case. 
 
A similar behaviour is observed on a larger scale in the right panel of  Fig.\ref{Fig4b}, 
where $\A_1$ passes through a minimum and then grows toward the past, while the other two 
amplitudes oscillate. Subsequently, $\A_3$ passes through a maximum and decreases 
toward the past, with the remaining two amplitudes continuing to oscillate.

\begin{figure}[b]
\hbox to \linewidth{ \hss

	
				\resizebox{11.5cm}{6.5cm}{\includegraphics{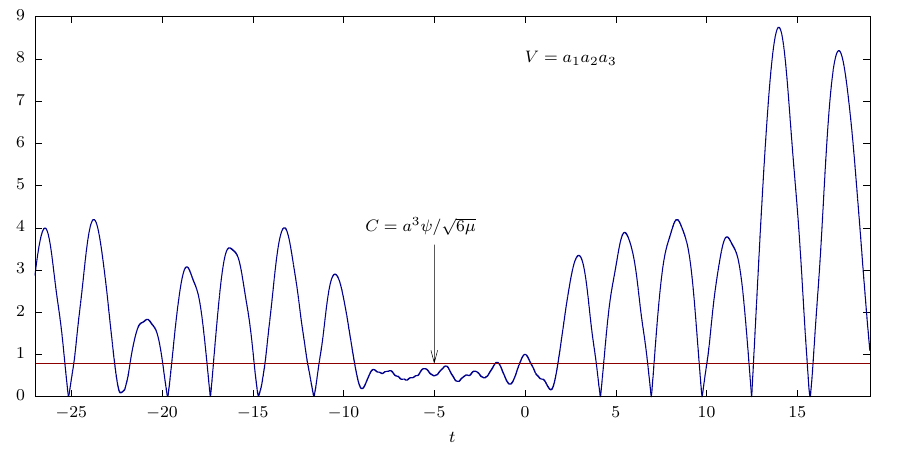}}

\hspace{1mm}
\hss}
\caption{Same solution as in Fig.\ref{Fig4b}, extended to positive values of $t$. 
The volume function  $V(t)$ exhibits  a seemingly random profile, 
quasi-periodically approaching zero but never reaching it. }
 \label{Fig5}
\end{figure}

As none of the amplitudes $\A_k(t)$ vanish, the curvature remains bounded and  the 
manifold is non-singular. The solution in the negative $t$ region can be extended to 
positive $t$, as  shown in Fig.\ref{Fig5}. 
Although $\dot{\A}(0)=0$,  the solution is not symmetric under $t\to -t$, since
the initial values of the anisotropies, $\dot{\beta}_\pm(0)$, do not vanish. 
It appears  that the range of $t$ extends from $-\infty$ to $\infty$, 
although we cannot immediately prove this.  
In principle,  it is not excluded that integrating far enough could lead to 
$V(t)=\A_1 \A_2 \A_3$ reaching zero at a minimum, producing a curvature singularity. 
However, our numerical results do not show this -- the integration terminates only because 
numerical errors accumulate and the constraint begins to be violated.

Qualitatively, it is the repulsive character of the phantom scalar field that prevents the 
system from reaching a singularity.

Therefore, the solution  shown in Fig.\ref{Fig5}  appears to correspond to a completely
regular, geodesically complete manifold. The distribution of peaks in   Fig.\ref{Fig5} appears 
entirely random; in fact, even a very  small change in the initial values parameters 
$\gamma$, $C$ completely reshuffles the positions and amplitudes of the peaks.  
This behaviour strongly suggests that the solution is chaotic.

As a result, we obtain an anisotropic universe whose three scale functions $\A_k(t)$ 
undergo an apparently infinite sequence of oscillation cycles distributed in a seemingly random way. 
These functions periodically approach zero (with random periods)   but never actually  
reach it, instead bouncing back. Consequently, the universe repeatedly  enters a strong-gravity regime 
without developing a curvature singularity. The size of the universe is also 
bounded from above, resulting in an infinite sequence of bounces. 
To the best of our knowledge, solutions of this type have not been described previously in the literature.

Since the phantom scalar field violates the null energy condition and provides a repulsive effect, 
the appearance of bounce solutions is, in some sense, natural. Such bounces already arise in the 
isotropic limit when matter is included --  for example a radiation with 
energy density $A^2/\A^4$.  In this case, Eq.\eqref{K00o} becomes 
\be
\frac{\dot{\A}^2}{\A^2}+\frac{1}{\A^2}=\frac{A^2}{\A^4}-\frac{C^2}{\A^6}, 
\ee
whose solution $\A(t)$ oscillates between the maximal and minimal values $\A_\pm$  of the universe size,  
given by $\A_\pm^2=(A^2\pm \sqrt{A^4-4C^2})/2$. It is less obvious, however,  that anisotropies can 
mimic the role of matter,  leading to the anisotropic bounces described above.

In summary, the energy of a minimally coupled massless  scalar field 
falls off as $\propto 1/\A^6$, just like the anisotropy contribution. As a result, 
the scalar is able to influence the chaotic dynamics in the Bianchi IX case: 
an ordinary scalar with positive energy removes the chaos, permitting only 
a finite number of oscillations as the singularity is approached. 
In contrast,
a phantom scalar appears to  eliminate  the singularity entirely, 
producing a chaotic  sequence of regular bounces.

\section{Anisotropic cosmologies with $K$-essence}

Let us now consider the case where the scalar field contribution is different from  $\propto 1/\A^6$. 
We take a $K$-essence theory with index $n\neq 2$, still assuming the minimal coupling 
($\alpha=0$). Then, the initial value constraint \eqref{K1} becomes 
\be                    \label{K00oo2}
 \frac{\dot{\A}^2}{\A^2}+\frac{\cal K}{\A^2}=\dot{\beta}_{+}^2+\dot{\beta}_{-}^2+(n-1)\frac{\epsilon}{6\mu}\,\psi^n ,
 \ee
 while Eq.\eqref{K4} yields 
 \be
 \A^3 \psi^{n-1}=const.~~~\Rightarrow~~~
 \psi=(6\mu)^{1/n}\,\frac{C}{\A^{3/(n-1)}}.
 \ee
 The constraint then reduces to 
 \be                    \label{K00oo3}
 \frac{\dot{\A}^2}{\A^2}+\frac{\cal K}{\A^2}=\dot{\beta}_{+}^2+\dot{\beta}_{-}^2
 +\epsilon\,\,(n-1)\, \frac{C^n}{\A^\gamma},~~~~\gamma=\frac{3n}{n-1}.
 \ee
 If $n>2$, then $\gamma<6$, so the scalar field contribution is $o(1/r^6)$ and 
 subdominant compared to the anisotropy contribution during Kasner cycles,
  which scales as 
 $\propto~1/\A^6$. According to our conjecture,
 the chaotic BKL-type behaviour is therefore expected in this case. Conversely,
 if $n<2$, then $\gamma>6$, and the scalar field 
 contribution dominates near singularity, 
 which should modify the  BKL picture. 
 
 \begin{figure}[t]
\hbox to \linewidth{ \hss
				\resizebox{6.3cm}{5cm}{\includegraphics{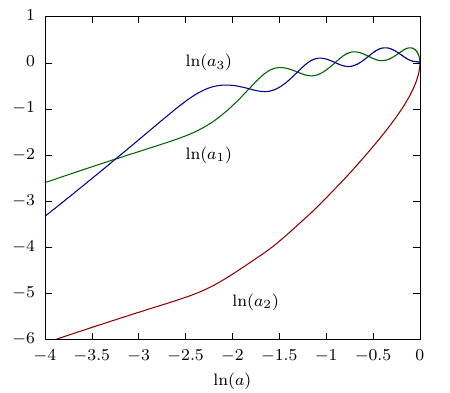}}
				\resizebox{6.3cm}{5cm}{\includegraphics{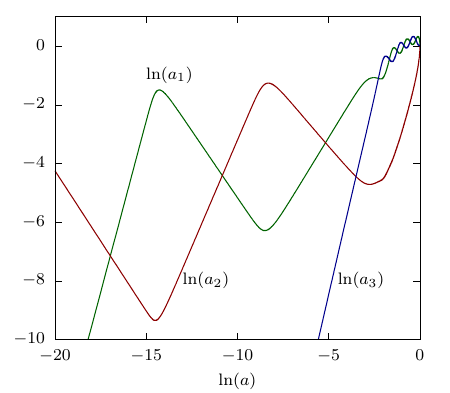}}	

\hspace{1mm}
\hss}
\caption{Bianchi IX solutions with $K$-essence ($\epsilon=1$) for the initial data \eqref{ini4}
with $\gamma=3$ and $C=0.123$, and for different values of the $K$-essence index $n$. Left: $n=1.9<2$; 
in this case, the scalar field contribution to \eqref{K00oo3} is dominant,  and 
the oscillations quickly stop, similar to the behaviour shown in Fig.\ref{Fig4a}. Right: $n=2.3>2$; here, 
the scalar contribution is subdominant, and the solution closely resembles the vacuum case shown in Fig.\ref{Fig0}.
}
 \label{Fig5a}
\end{figure}

 To verify these expectations, we use the initial values similar to those 
 in \eqref{ini3}:
 \be              \label{ini4}
 \A(0)&=&1,~~~\beta_\pm(0)=0,~~~~
 \dot{\beta}_{+}(0)={\cal B}\,\sin\gamma,~~~~\dot{\beta}_{-}(0)={\cal B}\cos\gamma, \nn \\
 \dot{\A}(0)&=&0,~~~~~~~\dot{\phi}=(6\mu)^{1/n} C,~~~~~~{\cal B}=\sqrt{1-\epsilon\, C^n}. 
 \ee
 Starting from these initial values, we integrate the equations 
 assuming $\epsilon=1$. Fig.\ref{Fig5a} shows the solutions for 
 $\gamma=3$ and $C=0.123$ for two different values of the $K$-essence index $n$. 
 For  $n=1.9<2$ (left panel), the scalar field 
dominates and the oscillations quickly stop, similarly to the  $n=2$ solution 
in Fig.\ref{Fig4a}. On the other hand, for $n=2.3>2$ (right panel), 
the scalar contribution is subdominant,  and the solution closely resembles 
 the vacuum case in Fig.\ref{Fig0}. 
This confirms the expectations outlined above.

\begin{figure}[t]
\hbox to \linewidth{ \hss	
				\resizebox{6.3cm}{5cm}{\includegraphics{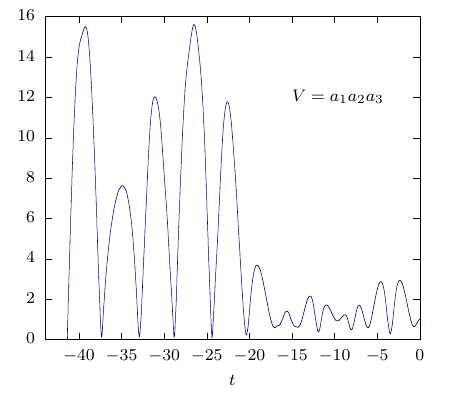}}
				\resizebox{6.3cm}{5cm}{\includegraphics{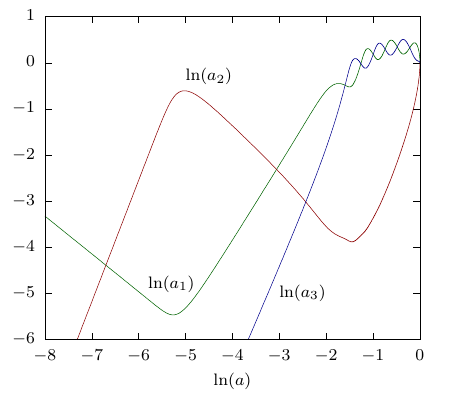}}	

\hspace{1mm}
\hss}
\caption{Bianchi IX solutions with phantom $K$-essence ($\epsilon=-1$) 
for the parameter values $\gamma=3$, $C=0.774$, and  for different values of the  $K$-essence  index $n$. 
Left: $n=1.5<2$; the scalar field dominates and provides a repulsion 
from the singularity, producing an apparently infinite chain of bounces, similar  to 
Fig.\ref{Fig4b}. 
Right: $n=3$; here, the scalar contribution is subdominant, 
and the solution closely resembles  the BKL case shown in Fig.\ref{Fig0}. 
}
 \label{Fig5b}
\end{figure}

Let us now consider the phantom case, $\epsilon=-1$. 
Fig.\ref{Fig5b} shows the solutions for 
 $\gamma=3$ and $C=0.774$.
For $n=1.5<2$ (left panel), the scalar field dominates the dynamics 
and provides a repulsion 
from singularity, producing an apparently infinite and chaotic chain of bounces, similar 
to the $n=2$ case in Fig.\ref{Fig4b}. 
For $n=3>2$ (right panel), the scalar field contribution is subdominant, and the 
solution closely resembles the vacuum BKL case in Fig.\ref{Fig0}.
Thus, the presence of a singularity depends on whether the scalar field is dominant or subdominant, 
but in all cases the dynamics remains chaotic.

Summarizing, for $\epsilon=1$ the scalar field has positive energy. 
If its contribution is subdominant, $o(1/\A^6)$, the solutions remain chaotic, 
whereas if it scales as $1/\A^6$ or faster, chaos is eliminated, confirming our conjecture. 
In contrast, for $\epsilon=-1$ the scalar energy is negative, and the solutions are always chaotic:
they develop a singularity when the scalar energy is subdominant ($o(1/\A^6)$) 
and remain regular otherwise. 

\section{Non-minimal coupling}
\setcounter{equation}{0}

Let us finally switch on the non-minimal coupling by setting 
$\alpha\neq 0$ in Eqs.\eqref{K1}-\eqref{K4}, 
while keeping the $K$-essence index $n=2$. 
The main difference from the $\alpha=0$ 
case is that, instead of $\psi=\sqrt{6\mu}C/\A^3$, Eq.\eqref{K4} now gives
\be
\psi=\frac{\sqrt{6\mu}\,C}{\A^3\,[1 - (y+\omega)/\zeta ]},
\ee
where 
\be
\zeta=\frac{\epsilon}{3\alpha},~~~~~~
y=\frac{\dot{\A}^2}{\A^2}-\dot{\beta}_{+}^2-\dot{\beta}_{-}^2,~~~~~
\omega =\frac{\cal K}{\A^2}.
\ee
Inserting this to the constraint \eqref{K1}, the result can be 
represented in the form 
\be            \label{aC}
\A^6=
\frac{\epsilon\zeta\,C^2\,(\zeta-3 y-\omega)}{(\zeta-y-\omega)^2
(y+\omega)}\,.
\ee
It is straightforward to verify that in the $\A\to 0$ limit this equation admits only one solution, 
\be
\A\to 0,~~~~~~y\to \frac{\zeta-\omega}{3},
\ee
so that 
\be                 \label{y}
y=\frac{\epsilon}{9\alpha}-\frac{\cal K}{3\A^2}+\ldots,~~~~~~~~~
\psi=\frac{\sqrt{6\mu}\,3 C}{2\A^3\, [1-{\cal K} /(\zeta\A ^2)]}+\ldots, 
\ee
where the dots denote subleading terms.

Let us now turn to the anisotropy equations \eqref{K3}. 
Neglecting their right-hand side parts  containing ${\cal K}$ gives
\be                \label{b}
\dot{\beta}_\pm=\frac{const.}{\sigma_{+} \A^3}=\frac{const.}{(2\mu+\alpha\,\psi^2)\, \A^3}. 
\ee
For $\alpha=0$, this equation  reduces to $\dot{\beta}_\pm \propto 1/\A^3$,
which is precisely the Kasner behaviour of the Bianchi~I solutions.
In the Bianchi~IX case, this corresponds to the oscillatory regime,
which can be approximated as a sequence of Kasner cycles.
Within each cycle, the ${\cal K}$ term in the equations can be neglected,
becoming important only at the transition between successive cycles.

If $\alpha\neq 0$, then in the Bianchi~I case 
Eqs.\eqref{y} and \eqref{b} give 
\be
{\cal K}=0:~~~~~
y\approx \frac{\epsilon}{9\alpha},~~~~~
\psi\propto \frac{1}{\A^3}~~~\To~~~~~\dot{\beta}_\pm\propto \A^3.
\ee
Therefore, as $\A\to 0$, the scalar field $\psi$ diverges while the anisotropies vanish.
This effect,  known as anisotropy screening,  was first discovered in \cite{Starobinsky:2019xdp}.

However, in  the Bianchi IX case, one obtains from 
\eqref{y} and \eqref{b}:
\be
{\cal K}\neq 0:~~~~~
y\propto \frac{\cal K}{\A^2},~~~~~
\psi\propto \frac{1}{{\cal K}\A}~~~\To~~~~~
\dot{\beta}_\pm\propto \frac{{\cal K}^2}{\A}\,,
\ee
therefore,  the anisotropy screening is destroyed  by the spatial curvature \cite{Starobinsky:2019xdp}. 

\begin{figure}[b]
\hbox to \linewidth{ \hss

				\resizebox{6.3cm}{5cm}{\includegraphics{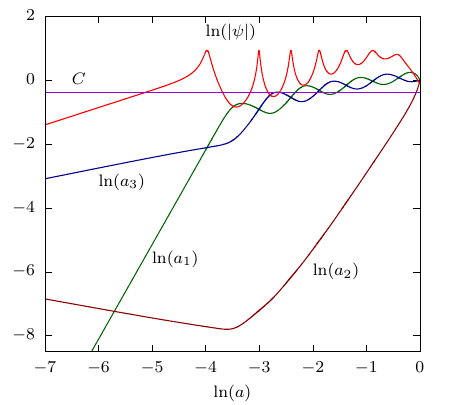}}
				\resizebox{6.3cm}{5cm}{\includegraphics{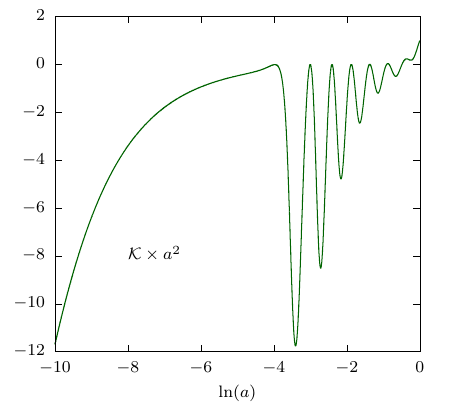}}	

\hspace{1mm}
\hss}
\caption{Bianchi IX solutions with a non-minimally coupled 
scalar  field for   $\alpha=-0.1$, $\epsilon=1$, $n=2$, and initial conditions 
$\A(0)=1$, $\beta_\pm(0) =0$, $\dot{\beta}_{+}(0)=\sin(3)$, 
 $\dot{\beta}_{-}=\cos(3)(0)$, and $\psi(0)=-0.9$.  
Left: the amplitudes $\A_k$, the scalar field $\psi$, and the scalar charge $C$ defined by \eqref{K4}. 
Right: The scalar curvature ${\cal K}\times \A^2$. 
}
 \label{Fig6}
\end{figure}

Using the definition of $y$ yields 
\be
\frac{\dot{\A}^2}{\A^2}=\dot{\beta}_{+}^2+\dot{\beta}_{-}^2+y.
\ee
This is actually the constraint equation, where the right-hand side contains both the 
anisotropy contribution and the scalar field contribution encoded in $y$. 
Therefore, $y$ can be identified with  the scalar field energy.
The solution is expected to be chaotic if the anisotropy 
term dominates. We therefore consider the ratio 
\be \label{Ko}
\frac{ \dot{\beta}_{+}^2+\dot{\beta}_{-}^2}{y}\propto \A^6\,~~~\text{if}~~~{\cal K}=0;~~~~~~~~~~~
\frac{ \dot{\beta}_{+}^2+\dot{\beta}_{-}^2}{y}\propto {\cal K}^3~~~\text{if}~~~{\cal K}\neq0.
\ee
Recall that ${\cal K}$ takes values in the interval
$(-\infty,1]$. If ${\cal K}$ is close to zero,
the first estimate in \eqref{K} applies. However,  as shown in Fig.\ref{Fig6} (right panel),
${\cal K}$ is typically large and negative, in which case 
the second estimate in \eqref{K} applies.
Therefore, for most of the evolution 
the anisotropy contribution dominates, and the solution is expected 
to be chaotic. 

These expectations are confirmed by numerical simulations. Fig.\ref{Fig6} (left panel) displays a solution 
exhibiting the characteristic BKL behaviour. The scalar curvature ${\cal K}$  periodically 
approaches zero but remains negative for most of the evolution, typically growing at least as fast as
$1/\A^2$. 

Solutions with  $\alpha\neq 0$ do not exhibit the same strong dependence on $\epsilon$  
as for $\alpha=0$. Solutions similar to that  shown in Fig.\ref{Fig6} exist for both signs of 
$\epsilon$ and $\alpha$. This can be attributed to the fact that, although in the non-minimally 
coupled case the scalar energy is not strictly positive, it is mainly positive near the singularity, where 
the estimate \eqref{y} applies, hence the energy is $\propto y\approx -{\cal K}/(3\A^2)>0$, 
because ${\cal K}$ is typically large and negative. 

The theory with $\alpha\neq 0$ also admits isotropic bounces whose size oscillates  between
 finite values $\A_{\pm}$ (see also \cite{Sushkov:2025udy}).  
 They can be obtained by solving Eq.\eqref{aC} with $\omega=1/\A^2$, 
 which yields $y(a)$, and this should be positive, which is indeed the case  for  
 $\A\in(\A_{-},\A_{+})$. 
 Such solutions  can be extended to the 
 anisotropic case, but no chaotic behaviour is observed.

\section{Summary}

We analyzed above how a scalar field affects the chaotic behaviour of 
homogeneous and anisotropic Bianchi IX cosmologies. 
Most of the time, we considered the scalar field
of the simplest shift-symmetric $K$-essence type, defined by 
$
X=\epsilon\, \dot{\phi}^n,
$
with $\epsilon=1$ or $\epsilon=-1$. 
It is known that such a field removes chaos 
if $n=2$  \cite{Belinskii:1973}, provided that $\epsilon=+1$. The spacetime singularity 
is then approached smoothly. 
We have verified  that chaos disappears also if $n<2$, because in this case the field energy 
$\propto 1/\A^{\gamma}$ with
 $\gamma>6$ dominates over the anisotropy contribution. However, for  $n>2$, the 
 scalar field contribution is subdominant, and the solutions exhibit  the usual BKL-type chaotic behaviour. 
 
 The BKL-type chaos also persists when a non-minimal coupling is included \cite{Starobinsky:2019xdp},
 and we have confirmed  that in this case as well the scalar field contribution remains subdominant.

 The BKL-type behaviour  also persists for the phantom scalar field corresponding to $\epsilon=-1$
 if $n>2$. However,  for $n\leq 2$ we obtain a completely new phenomenon: chaos remains 
 but the spacetime singularity disappears. In this regime, the three scale factors $\A_k(t)$ oscillate seemingly 
 indefinitely, never reaching zero, and the 3-volume $V=\A_1\A_2\A_3$ exhibits a chaotic sequence 
 of minima and maxima. The universe thus undergoes an infinite sequence of anisotropic bounces.

It is not immediately clear whether this finding has direct physical applications, even though theories 
with non-positive energy have become increasingly popular in recent years. 
Nevertheless, these new chaotic solutions are interesting from a theoretical perspective. 
For example, one could study their topological entropy \cite{Kamenshchik:1998ix} or investigate 
other characteristics of their chaotic behaviour.

\begin{acknowledgement}
It is a pleasure to thank Sasha Kamenshchik for valuable discussions. 
\end{acknowledgement}




\begin{thebibliography}{10}

\bibitem{Starobinsky:2016kua}
A.~A. Starobinsky, S.~V. Sushkov, and M.~S. Volkov, {\it {The screening
  Horndeski cosmologies}},  {\sl JCAP} {\bf 1606} (2016), no.~06 007,
  [\href{http://dx.doi.org/10.1088/1475-7516/2016/06/007}{{\sf
  doi:10.1088/1475-7516/2016/06/007}}].

\bibitem{Starobinsky:2019xdp}
A.~A. Starobinsky, S.~V. Sushkov, and M.~S. Volkov, {\it {Anisotropy screening
  in Horndeski cosmologies}},  {\sl Phys. Rev. D} {\bf 101} (2020), no.~6
  064039, [\href{http://arxiv.org/abs/1912.12320}{{\sf arXiv:1912.12320}}],
  [\href{http://dx.doi.org/10.1103/PhysRevD.101.064039}{{\sf
  doi:10.1103/PhysRevD.101.064039}}].

\bibitem{Galeev:2021xit}
R.~Galeev, R.~Muharlyamov, A.~A. Starobinsky, S.~V. Sushkov, and M.~S. Volkov,
  {\it {Anisotropic cosmological models in Horndeski gravity}},  {\sl Phys.
  Rev. D} {\bf 103} (2021), no.~10 104015,
  [\href{http://arxiv.org/abs/2102.10981}{{\sf arXiv:2102.10981}}],
  [\href{http://dx.doi.org/10.1103/PhysRevD.103.104015}{{\sf
  doi:10.1103/PhysRevD.103.104015}}].

\bibitem{Belinskii:1972}
V.~Belinskii, E.~Lifshitz, and I.~Khalatnikov, {\it {Construction of a general
  cosmological solution of the Einstein equation with a time singularity}},
  {\sl Sov.JETP} {\bf 35} (1972) 838--841.

\bibitem{Misner:1969hg}
C.~W. Misner, {\it {Mixmaster universe}},  {\sl Phys. Rev. Lett.} {\bf 22}
  (1969) 1071--1074, [\href{http://dx.doi.org/10.1103/PhysRevLett.22.1071}{{\sf
  doi:10.1103/PhysRevLett.22.1071}}].

\bibitem{Goldstein:2025qxq}
P.~P. Goldstein, {\it {Some exact results on the Belinski-Khalatnikov-Lifshitz
  scenario}},  \href{http://arxiv.org/abs/2505.15541}{{\sf arXiv:2505.15541}}.

\bibitem{Belinskii:1973}
V.~Belinskii, E.M., and I.~Khalatnikov, {\it {Effect of scalar and vector
  fields on the nature of the cosmological singularity }},  {\sl Sov.JETP} {\bf
  36} (1972) 591--597.

\bibitem{Kamenshchik:1998ix}
A.~Y. Kamenshchik, I.~M. Khalatnikov, S.~V. Savchenko, and A.~V. Toporensky,
  {\it {Topological entropy for some isotropic cosmological models}},  {\sl
  Phys. Rev. D} {\bf 59} (1999) 123516,
  [\href{http://arxiv.org/abs/gr-qc/9809048}{{\sf arXiv:gr-qc/9809048}}],
  [\href{http://dx.doi.org/10.1103/PhysRevD.59.123516}{{\sf
  doi:10.1103/PhysRevD.59.123516}}].

\bibitem{Easson:2024fzn}
D.~A. Easson and J.~E. Lesnefsky, {\it {Eternal Universes}},
  \href{http://arxiv.org/abs/2404.03016}{{\sf arXiv:2404.03016}},
  \href{http://dx.doi.org/10.1103/5mhz-m8bg}{{\sf doi:10.1103/5mhz-m8bg}}.

\bibitem{Sushkov:2025udy}
S.~V. Sushkov and R.~G. Galeev, {\it {Singular bounce in the theory of gravity
  with nonminimal derivative coupling}},  {\sl Phys. Rev. D} {\bf 111} (2025),
  no.~8 083554, [\href{http://arxiv.org/abs/2502.05786}{{\sf
  arXiv:2502.05786}}],
  [\href{http://dx.doi.org/10.1103/PhysRevD.111.083554}{{\sf
  doi:10.1103/PhysRevD.111.083554}}].

\end{thebibliography}




\end{document}